\documentclass[prd,twocolumn,showpacs,amsmath,amssymb]{revtex4}

\usepackage{graphicx}
\usepackage{dcolumn}
\usepackage{bm}

\begin{document}




\hspace{5.2in} \mbox{FERMILAB-PUB-06/327-E}
\title{Measurement of the CP-violation parameter
of $B^0$ mixing and decay with
$p \bar{p} \rightarrow \mu \mu X$ data
}


%
\author{                                                                      
V.M.~Abazov,$^{35}$                                                           
B.~Abbott,$^{75}$                                                             
M.~Abolins,$^{65}$                                                            
B.S.~Acharya,$^{28}$                                                          
M.~Adams,$^{51}$                                                              
T.~Adams,$^{49}$                                                              
M.~Agelou,$^{17}$                                                             
E.~Aguilo,$^{5}$                                                              
S.H.~Ahn,$^{30}$                                                              
M.~Ahsan,$^{59}$
% ...
\\                                                                            
\vskip 0.30cm                                                                 
\centerline{(D\O\ Collaboration)}                                             
\vskip 0.30cm                                                                 
}                                                                             
\affiliation{                                                                 
\centerline{$^{1}$Universidad de Buenos Aires, Buenos Aires, Argentina}       
\centerline{$^{2}$LAFEX, Centro Brasileiro de Pesquisas F{\'\i}sicas,         
                  Rio de Janeiro, Brazil}                                     
\centerline{$^{3}$Universidade do Estado do Rio de Janeiro,                   
                  Rio de Janeiro, Brazil}                                     
\centerline{$^{4}$Instituto de F\'{\i}sica Te\'orica, Universidade            
                  Estadual Paulista, S\~ao Paulo, Brazil}                     
\centerline{$^{5}$University of Alberta, Edmonton, Alberta, Canada,           
                  Simon Fraser University, Burnaby, British Columbia, Canada,}
\centerline{York University, Toronto, Ontario, Canada, and                    
                  McGill University, Montreal, Quebec, Canada}                
\centerline{$^{6}$University of Science and Technology of China, Hefei,       
                  People's Republic of China}                                 
\centerline{$^{7}$Universidad de los Andes, Bogot\'{a}, Colombia}             
\centerline{$^{8}$Center for Particle Physics, Charles University,            
                  Prague, Czech Republic}                                     
\centerline{$^{9}$Czech Technical University, Prague, Czech Republic}         
\centerline{$^{10}$Center for Particle Physics, Institute of Physics,         
                   Academy of Sciences of the Czech Republic,                 
                   Prague, Czech Republic}                                    
\centerline{$^{11}$Universidad San Francisco de Quito, Quito, Ecuador}        
\centerline{$^{12}$Laboratoire de Physique Corpusculaire, IN2P3-CNRS,         
                   Universit\'e Blaise Pascal, Clermont-Ferrand, France}      
\centerline{$^{13}$Laboratoire de Physique Subatomique et de Cosmologie,      
                   IN2P3-CNRS, Universite de Grenoble 1, Grenoble, France}    
\centerline{$^{14}$CPPM, IN2P3-CNRS, Universit\'e de la M\'editerran\'ee,     
                   Marseille, France}                                         
\centerline{$^{15}$IN2P3-CNRS, Laboratoire de l'Acc\'el\'erateur              
                   Lin\'eaire, Orsay, France}                                 
\centerline{$^{16}$LPNHE, IN2P3-CNRS, Universit\'es Paris VI and VII,         
                   Paris, France}                                             
\centerline{$^{17}$DAPNIA/Service de Physique des Particules, CEA, Saclay,    
                   France}                                                    
\centerline{$^{18}$IPHC, IN2P3-CNRS, Universit\'e Louis Pasteur, Strasbourg,  
                    France, and Universit\'e de Haute Alsace,                 
                    Mulhouse, France}                                         
\centerline{$^{19}$Institut de Physique Nucl\'eaire de Lyon, IN2P3-CNRS,      
                   Universit\'e Claude Bernard, Villeurbanne, France}         
\centerline{$^{20}$III. Physikalisches Institut A, RWTH Aachen,               
                   Aachen, Germany}                                           
\centerline{$^{21}$Physikalisches Institut, Universit{\"a}t Bonn,             
                   Bonn, Germany}                                             
\centerline{$^{22}$Physikalisches Institut, Universit{\"a}t Freiburg,         
                   Freiburg, Germany}                                         
\centerline{$^{23}$Institut f{\"u}r Physik, Universit{\"a}t Mainz,            
                   Mainz, Germany}                                            
\centerline{$^{24}$Ludwig-Maximilians-Universit{\"a}t M{\"u}nchen,            
                   M{\"u}nchen, Germany}                                      
\centerline{$^{25}$Fachbereich Physik, University of Wuppertal,               
                   Wuppertal, Germany}                                        
\centerline{$^{26}$Panjab University, Chandigarh, India}                      
\centerline{$^{27}$Delhi University, Delhi, India}                            
\centerline{$^{28}$Tata Institute of Fundamental Research, Mumbai, India}     
\centerline{$^{29}$University College Dublin, Dublin, Ireland}                
\centerline{$^{30}$Korea Detector Laboratory, Korea University,               
                   Seoul, Korea}                                              
\centerline{$^{31}$SungKyunKwan University, Suwon, Korea}                     
\centerline{$^{32}$CINVESTAV, Mexico City, Mexico}                            
\centerline{$^{33}$FOM-Institute NIKHEF and University of                     
                   Amsterdam/NIKHEF, Amsterdam, The Netherlands}              
\centerline{$^{34}$Radboud University Nijmegen/NIKHEF, Nijmegen, The          
                  Netherlands}                                                
\centerline{$^{35}$Joint Institute for Nuclear Research, Dubna, Russia}       
\centerline{$^{36}$Institute for Theoretical and Experimental Physics,        
                   Moscow, Russia}                                            
\centerline{$^{37}$Moscow State University, Moscow, Russia}                   
\centerline{$^{38}$Institute for High Energy Physics, Protvino, Russia}       
\centerline{$^{39}$Petersburg Nuclear Physics Institute,                      
                   St. Petersburg, Russia}                                    
\centerline{$^{40}$Lund University, Lund, Sweden, Royal Institute of          
                   Technology and Stockholm University, Stockholm,            
                   Sweden, and}                                               
\centerline{Uppsala University, Uppsala, Sweden}                              
\centerline{$^{41}$Physik Institut der Universit{\"a}t Z{\"u}rich,            
                   Z{\"u}rich, Switzerland}                                   
\centerline{$^{42}$Lancaster University, Lancaster, United Kingdom}           
\centerline{$^{43}$Imperial College, London, United Kingdom}                  
\centerline{$^{44}$University of Manchester, Manchester, United Kingdom}      
\centerline{$^{45}$University of Arizona, Tucson, Arizona 85721, USA}         
\centerline{$^{46}$Lawrence Berkeley National Laboratory and University of    
                   California, Berkeley, California 94720, USA}               
\centerline{$^{47}$California State University, Fresno, California 93740, USA}
\centerline{$^{48}$University of California, Riverside, California 92521, USA}
\centerline{$^{49}$Florida State University, Tallahassee, Florida 32306, USA} 
\centerline{$^{50}$Fermi National Accelerator Laboratory,                     
            Batavia, Illinois 60510, USA}                                     
\centerline{$^{51}$University of Illinois at Chicago,                         
            Chicago, Illinois 60607, USA}                                     
\centerline{$^{52}$Northern Illinois University, DeKalb, Illinois 60115, USA} 
\centerline{$^{53}$Northwestern University, Evanston, Illinois 60208, USA}    
\centerline{$^{54}$Indiana University, Bloomington, Indiana 47405, USA}       
\centerline{$^{55}$University of Notre Dame, Notre Dame, Indiana 46556, USA}  
\centerline{$^{56}$Purdue University Calumet, Hammond, Indiana 46323, USA}    
\centerline{$^{57}$Iowa State University, Ames, Iowa 50011, USA}              
\centerline{$^{58}$University of Kansas, Lawrence, Kansas 66045, USA}         
\centerline{$^{59}$Kansas State University, Manhattan, Kansas 66506, USA}     
\centerline{$^{60}$Louisiana Tech University, Ruston, Louisiana 71272, USA}   
\centerline{$^{61}$University of Maryland, College Park, Maryland 20742, USA} 
\centerline{$^{62}$Boston University, Boston, Massachusetts 02215, USA}       
\centerline{$^{63}$Northeastern University, Boston, Massachusetts 02115, USA} 
\centerline{$^{64}$University of Michigan, Ann Arbor, Michigan 48109, USA}    
\centerline{$^{65}$Michigan State University,                                 
            East Lansing, Michigan 48824, USA}                                
\centerline{$^{66}$University of Mississippi,                                 
            University, Mississippi 38677, USA}                               
\centerline{$^{67}$University of Nebraska, Lincoln, Nebraska 68588, USA}      
\centerline{$^{68}$Princeton University, Princeton, New Jersey 08544, USA}    
\centerline{$^{69}$State University of New York, Buffalo, New York 14260, USA}
\centerline{$^{70}$Columbia University, New York, New York 10027, USA}        
\centerline{$^{71}$University of Rochester, Rochester, New York 14627, USA}   
\centerline{$^{72}$State University of New York,                              
            Stony Brook, New York 11794, USA}                                 
\centerline{$^{73}$Brookhaven National Laboratory, Upton, New York 11973, USA}
\centerline{$^{74}$Langston University, Langston, Oklahoma 73050, USA}        
\centerline{$^{75}$University of Oklahoma, Norman, Oklahoma 73019, USA}       
\centerline{$^{76}$Oklahoma State University, Stillwater, Oklahoma 74078, USA}
\centerline{$^{77}$Brown University, Providence, Rhode Island 02912, USA}     
\centerline{$^{78}$University of Texas, Arlington, Texas 76019, USA}          
\centerline{$^{79}$Southern Methodist University, Dallas, Texas 75275, USA}   
\centerline{$^{80}$Rice University, Houston, Texas 77005, USA}                
\centerline{$^{81}$University of Virginia, Charlottesville,                   
            Virginia 22901, USA}                                              
\centerline{$^{82}$University of Washington, Seattle, Washington 98195, USA}  
}                                                                             

\date{September 7, 2006}
           
\begin{abstract}

\noindent
We measure the dimuon charge asymmetry $A$ in
$p \bar{p}$ collisions at a center of mass energy
$\sqrt{s} = 1960$ GeV. The data was
recorded with the D0 detector and corresponds to an
integrated luminosity of approximately 1.0 fb$^{-1}$.
Assuming that the asymmetry $A$ is due to
asymmetric $B^0 \leftrightarrow \bar{B}^0$ mixing and decay,
we extract the CP-violation parameter
of $B^0$ mixing and decay:
\begin{eqnarray*}
\frac{\Re{(\epsilon_{B^0})}}{1 + \left| \epsilon_{B^0} \right|^2} =
\frac{A_{B^0}}{4} =
-0.0023 \pm 0.0011\textrm{ (stat)} \pm 0.0008\textrm{ (syst)}.
\end{eqnarray*}
$A_{B^0}$ is the dimuon charge asymmetry from decays
of $B^0 \bar{B}^0$ pairs.
The general case, with CP violation in both $B^0$ and $B_s^0$
systems, is also considered.
Finally we obtain
the forward-backward asymmetry that quantifies the
tendency of $\mu^+$ to go in the proton direction and
$\mu^-$ to go in the anti-proton direction.
The results are consistent with the standard model
and constrain new physics.


\end{abstract}


\maketitle

\newpage

\section{Introduction}
We measure the dimuon charge asymmetry:
\begin{equation}
A = \frac{N^{++} - N^{--}}{N^{++} + N^{--}}
\label{o_defA}
\end{equation}
in $p \bar{p}$ collisions at a center of mass energy
$\sqrt{s} = 1960$ GeV.
$N^{++}$ ($N^{--}$) is the number of events with two
positive (negative) muon candidates passing selection cuts.
The data was
recorded with the D0 detector 
at the Fermilab Tevatron between 2002 and 2005.
The exposed
integrated luminosity is approximately 1.0 fb$^{-1}$.
Assuming that the asymmetry $A$  is due to
asymmetric $B^0 \leftrightarrow \bar{B}^0$ mixing and decay,
we extract the CP-violation parameter
of $B^0$ mixing and decay~\cite{pdg, buras, randall}:
\begin{equation}
\frac{\Re{(\epsilon_{B^0})}}{1 + \left| \epsilon_{B^0} \right|^2} = 
\Im{ \left\{ \frac{\Gamma_{12}}{4 M_{12}} \right\}} = \frac{A_{B^0}}{4}
\equiv f \cdot A.
\label{epsilon}
\end{equation}
$M_{12}$ ($\Gamma_{12}$) is the real
(imaginary) part of the transition matrix element of the Hamiltonian
corresponding to $(B^0, \bar{B}^0)$ mixing and decay.
Throughout this article we use
the Particle Data Group \cite{pdg} notation: 
$B^0 = d \bar{b}$, $B^0_s = s \bar{b}$.
$A_{B^0}$ is the dimuon charge asymmetry from direct-direct decays
of $B^0 \bar{B}^0$ (we define \textquotedblleft{direct decay}" 
as $b \rightarrow \mu^- X$,
and \textquotedblleft{sequential decay}" 
as $b \rightarrow c \rightarrow \mu^+ X$).
The dimuon charge asymmetry $A$ in Eq.~(\ref{epsilon}) excludes
events with a muon from $K^\pm$ decay.
Equation (\ref{epsilon}) defines the factor $f$, to be obtained below,
which accounts for other processes contributing to dimuon events.
As a sensitive cross check, we also measure the mean mixing probability
$\chi_0$ of $B \leftrightarrow \bar{B}$
hadrons, averaged over the mix of hadrons with a $b$ quark.
Finally we obtain
the forward-backward asymmetry that quantifies the
tendency of $\mu^+$ to go in the proton direction and
$\mu^-$ to go in the anti-proton direction.

The general case, with CP violation
in both $B^0$ and $B_s^0$ systems,
is considered in the last section of this article.
In this general case, the dimuon charge
asymmetry $A$ has contributions from both
$B^0$ and $B_s^0$.
Therefore, this measurement at the Fermilab Tevatron $p \bar{p}$
collider is complementary to similar measurements at
$B$ factories that are sensitive only to $A_{B^0}$,
not $A_{B_s^0}$.

The CP-violation parameter, defined in Eq.~(\ref{epsilon}), is sensitive to
several extensions of the standard model because new particles
may contribute to the box diagrams of $M_{12}$ \cite{randall,hewett}.
Reference \cite{randall} concludes that
\textquotedblleft{It is possible that the dilepton
asymmetry could be one of the first indications
of physics beyond the standard model}".

The D0 detector has an excellent muon system in 
Run II \cite{run2muon}, with 
large $(\eta, \phi)$ coverage, good scintillator-based triggering and
cosmic ray rejection, low punch-through rate, and precision tracking.
The muon is the particle with cleanest identification.
The like-sign dimuon channel is particularly clean: 
few processes contribute to it
and fewer still contribute to an asymmetry.
The D0 detector is well suited
for this precision measurement.

The outline of the paper is as follows.
The D0 detector is described in Section II.
In Section III we consider the event selection.
Physics and detector asymmetries are studied in Section IV.
The processes contributing to the asymmetry $A$ are
presented in Section V, and their weights are summarized in
Section VI.
The breakdown of systematic uncertainties of $A$
is discussed in Section VII.
Cross-checks are listed in Section VIII.
Final results are summarized in Section IX.

\section{The D0 detector}
The D0 detector consists of a magnetic central-tracking system,
comprised of a silicon microstrip tracker (SMT) and a central fiber
tracker (CFT), both located within a 2~T superconducting solenoidal
magnet~\cite{run2det}. The SMT has $\approx 800,000$ individual strips,
with typical pitch of $50-80$ $\mu$m, and a design optimized for
tracking and vertexing capability at pseudorapidities of $|\eta|<2.5$.
The system has a six-barrel longitudinal structure, each with a set
of four layers arranged axially around the beam pipe, and interspersed
with 16 radial disks. The CFT has eight thin coaxial barrels, each
supporting two doublets of overlapping scintillating fibers of 0.835~mm
diameter, one doublet being parallel to the collision axis, and the
other alternating by $\pm 3^{\circ}$ relative to the axis. Light signals
are transferred via clear fibers to solid-state photon counters (VLPC)
that have $\approx 80$\% quantum efficiency.

Central and forward preshower detectors located just outside of the
superconducting coil (in front of the calorimetry) are constructed of
several layers of extruded triangular scintillator strips that are read
out using wavelength-shifting fibers and VLPCs. The next layer of
detection involves three liquid-argon/uranium calorimeters: a central
section (CC) covering $|\eta|$ up to $\approx 1.1$, and two endcap
calorimeters (EC) that extend coverage to $|\eta|\approx 4.2$, all
housed in separate cryostats~\cite{run1det}. In addition to the preshower
detectors, scintillators between the CC and EC cryostats provide
sampling of developing showers at $1.1<|\eta|<1.4$.

A muon system~\cite{run2muon} is located beyond the calorimetry, and consists of a
layer of tracking detectors and scintillation trigger counters
before 1.8~T iron toroids, followed by two similar layers after
the toroids. Tracking at $|\eta|<1$ relies on 10~cm wide drift
tubes~\cite{run1det}, while 1~cm mini-drift tubes are used at
$1<|\eta|<2$.

A muon originating in a $p \bar{p}$ collision traverses
the silicon microstrip tracker and the scintillating
fiber tracker in the 2 T solenoidal magnetic field,
the calorimeter, layer \textsf{A} of the muon spectrometer,
the 1.8 T magnetized iron toroid, 
and layers \textsf{B} and \textsf{C} of the spectrometer.

Luminosity is measured using plastic scintillator arrays located in front
of the EC cryostats, covering $2.7 < |\eta| < 4.4$. 

Trigger and data acquisition systems are designed to accommodate
the high luminosities of Run II. Based on preliminary information from
tracking, calorimetry, and muon systems, the output of the first level
of the trigger is used to limit the rate for accepted events to
$\approx$ 2~kHz. At the next trigger stage, with more refined
information, the rate is reduced further to $\approx$ 1~kHz. These
first two levels of triggering rely mainly on hardware and firmware.
The third and final level of the trigger, with access to all the event
information, uses software algorithms and a computing farm, and reduces
the output rate to $\approx$ 50~Hz, which is written to tape.

The polarities of the toroid and solenoid magnetic fields are reversed
roughly every two weeks so that the four solenoid-toroid polarity
combinations are exposed to approximately the same
integrated luminosity. This allows cancellation of first-order
effects of the detector geometry. 

\section{Event selection}
Our standard cuts require global muons, i.e., 
local muon candidates
(reconstructed from hits in layers 
\textsf{A}, \textsf{B} and \textsf{C})
with a matching central track
(reconstructed from hits in the silicon and fiber trackers).
To reduce punch-through of hadrons
we only consider muons that traverse the
iron toroid. 
To select muons that emerge from the 
toroid with momentum $p \gtrsim 0.2$ GeV/$c$, we require 
$p_T > 4.2$ GeV/$c$ \textit{or} $|p_z| > 6.4$ GeV/$c$, where $p_T$ is the momentum
transverse to the beam measured by the central tracking system, 
and $p_z$ is the component of the
momentum in the direction of the proton beam.
We require at least two wire chamber hits in the \textsf{A} layer and
at least three wire chamber hits in layers \textsf{B} or \textsf{C}. 
We require local and global track fits with good $\chi^2$.
To reduce cosmic ray background, we require at least one 
scintillator hit associated with the muon to be within a time window 
of $\pm 5$~ns with respect to the expected time. 
To reduce muons from
$K^\pm$ and $\pi^\pm$ decay, we require $p_T > 3.0$ GeV/$c$.
The track is required to have a distance of closest approach to the
beam less than 0.3 cm.
We use the full pseudorapidity range $\left| \eta \right| < 2.2$.
We apply a cut of $p_T < 15.0$ GeV/$c$ 
to reduce the number of muons
reconstructed with wrong sign, and to reduce the
background from $W^\pm$ and $Z$ boson decay.
This list completes the single muon cuts.

The dimuon cuts are as follows.
We require that both muon candidates pass 
within 2.0 cm of each other in the direction along the beam line
at the point of closest approach to the beam.
To further reduce cosmic rays
and repeated reconstructions of the same track
(with different hits), 
we require the 3-dimensional 
opening angle between the
muons to be between $10^\circ$ and $170^\circ$. 
We also require that the two muons have different \textsf{A} layer
positions (by at least 5 cm), different 
local momentum vectors (by at least 0.2 GeV/$c$), 
and different central track momentum vectors (by at least 0.2 GeV/$c$).
This completes the set of standard cuts.

To avoid a bias due to mismatched central tracks 
(which are measured to be charge-asymmetric)
we use the local
muon charge, instead of the matching central track charge,
to obtain the asymmetries.
The positive charge asymmetry of central tracks is due to 
secondary particles that emerge from interactions with
detector material. The measured 
charge asymmetry of central tracks with $p_T > 3.0$ GeV/$c$ is
$(N^+ - N^-)/(N^+ + N^-) = 0.0049 \pm 0.0005$. 

The measurement of the dimuon charge asymmetry $A$
is based on \textit{ratios of muon counts}.
To minimize the statistical error we use all 
recorded events regardless of trigger.
If an event passes cuts, it is valid to accept it regardless
of trigger, since an event with opposite muon charges
can also be accepted by that trigger. 

\begin{figure*}
\begin{center}
\scalebox{0.9}
{\includegraphics{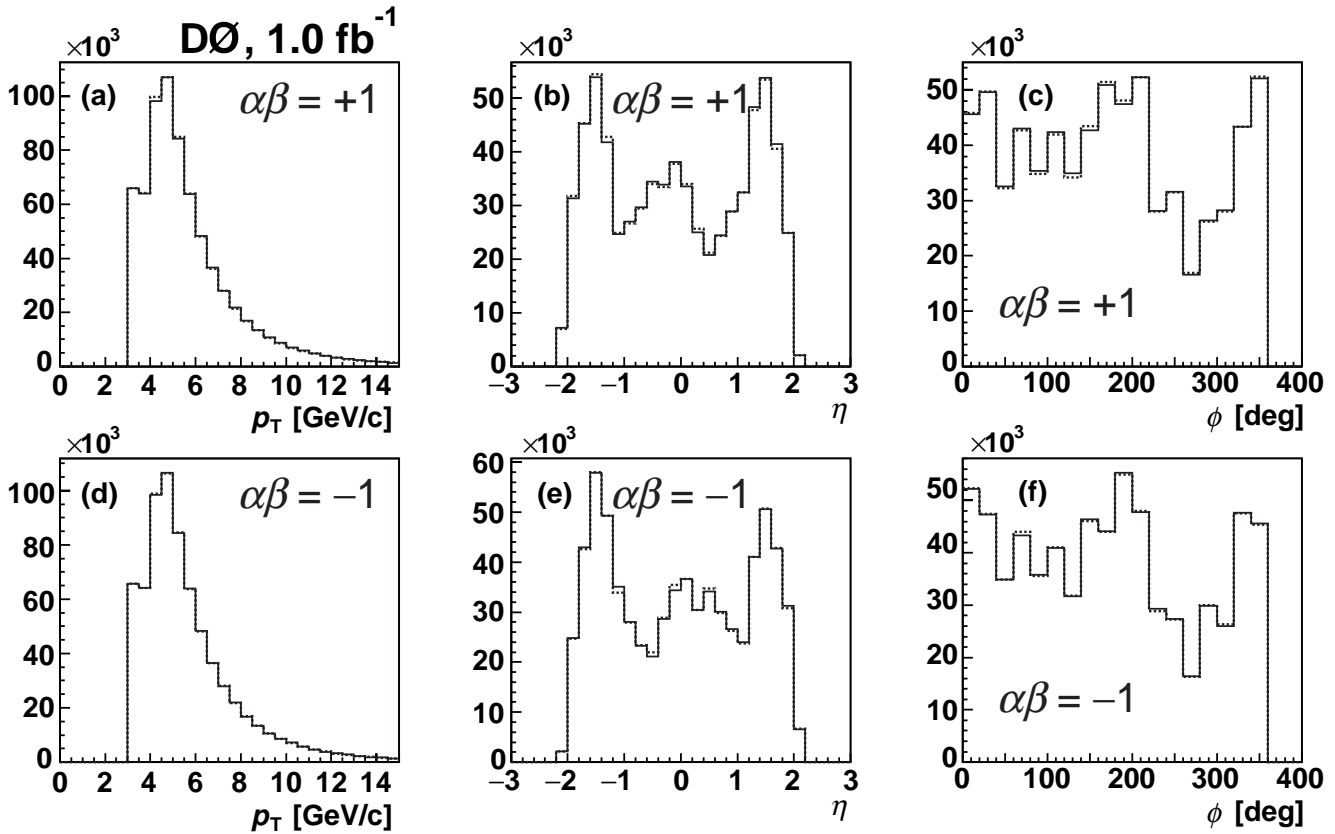}}
\caption{Distributions of single muon
 $p_T$ (a) and (d), $\eta$ (b) and (e), and $\phi$ (c) and (f), for events
with opposite toroid and solenoid polarities passing
standard single muon and dimuon cuts. 
The charge times toroid polarity, $\alpha \beta$, 
is indicated in the labels. 
Each plot superposes two histograms
with opposite charge and opposite polarities.
Positive (negative) charge corresponds to solid (dotted) lines
(that can generally not be  distinguished).
Histograms with negative toroid polarity are scaled
by the ratio of muon counts $e = 1.0300$, 
see Section IV.}
\label{pt_eta_phi_opp}
\end{center}
\end{figure*}

\begin{figure*}
\begin{center}
\scalebox{0.9}
{\includegraphics{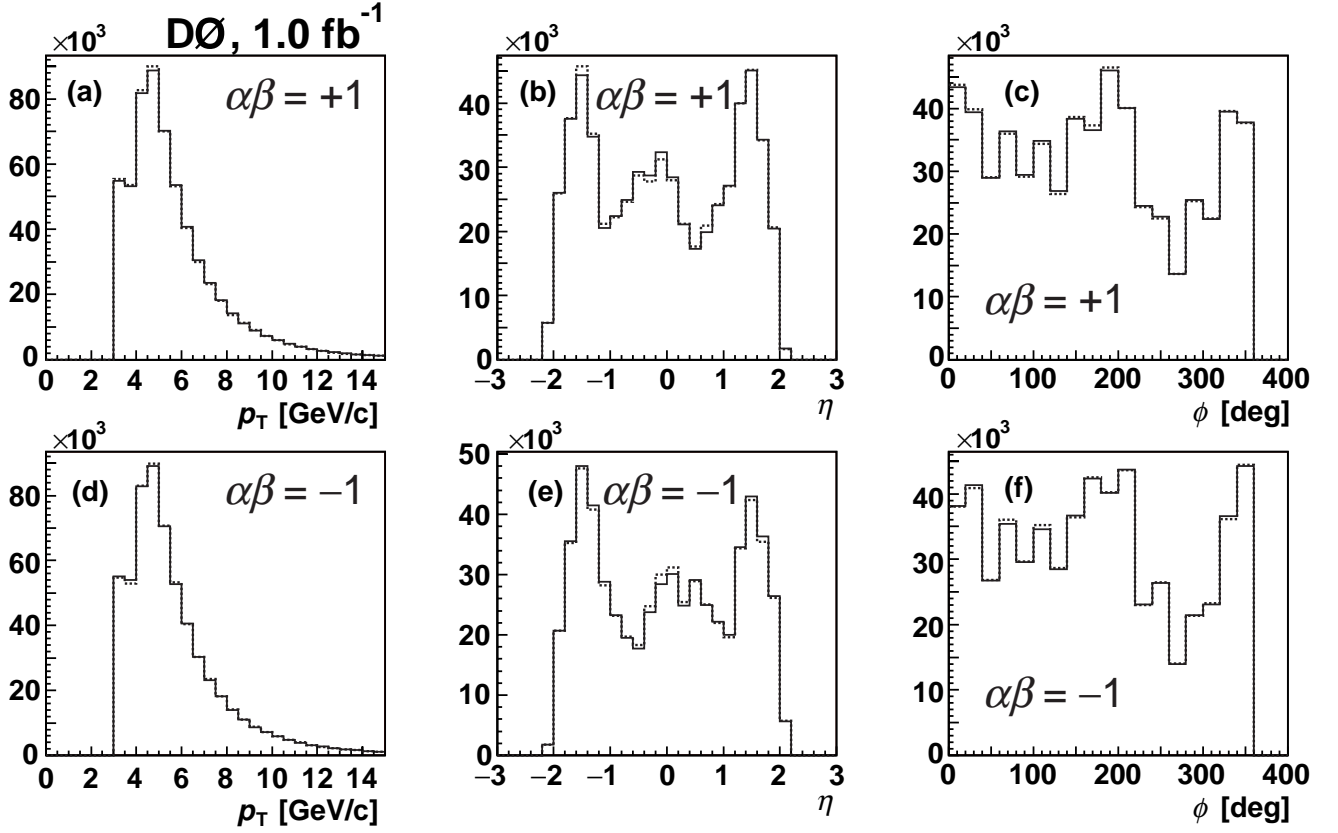}}
\caption{Same as Fig. \ref{pt_eta_phi_opp}, but for
equal toroid and solenoid polarities. Here $e = 0.9292$.}
\label{pt_eta_phi_same}
\end{center}
\end{figure*}

Histograms of $p_T$, $\eta$, and $\phi$ for standard 
cuts are shown in Figs.
\ref{pt_eta_phi_opp} and \ref{pt_eta_phi_same}.
Each plot superposes two histograms
with opposite charge and opposite polarities (that can
generally not be distinguished).

\begin{figure}
\begin{center}
\scalebox{0.59}
{\includegraphics{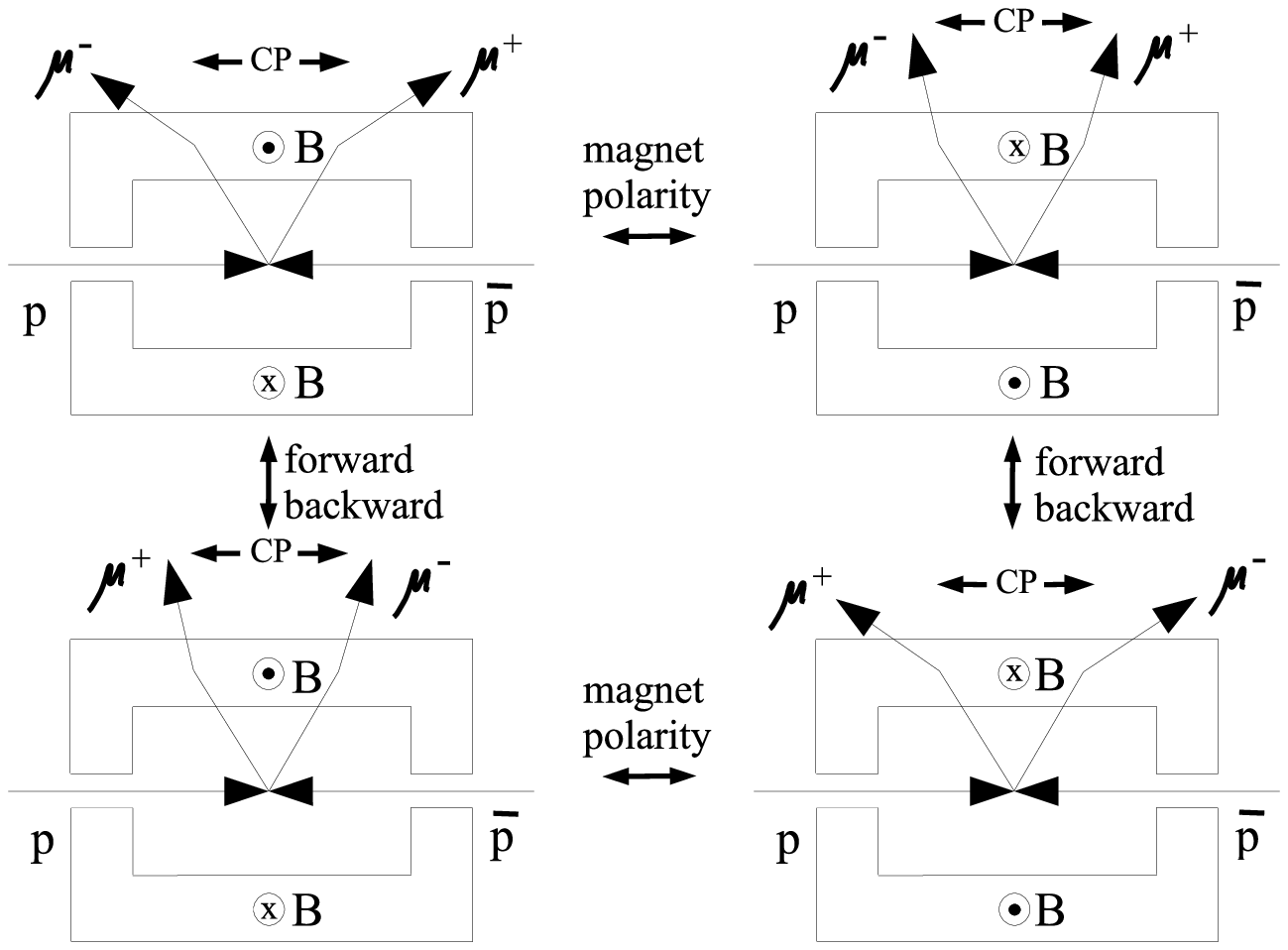}}
\caption{Schematic drawing of the magnetized iron toroids
of the D0 detector, and muon tracks related
by toroid polarity reversal, CP conjugation and
forward-backward reflection.}
\label{detector}
\end{center}
\end{figure}

\section{Asymmetries}
In this Section we study detector and physics effects that may alter
the charge asymmetry and therefore contribute to
corrections and systematic uncertainties. 
The muon detector is shown schematically in Fig. \ref{detector}.
How can the detector introduce a charge asymmetry?
It would have to have different acceptance $\times$
efficiency for positive and negative muons, i.e., 
for tracks bending north and south in the magnetized
iron toroid, see Fig. \ref{detector}. 
Such a difference may be due to an
offset of the mean beam spot, to mechanical asymmetries,
and to differences in wire chamber and scintillator efficiencies.
We find that the detector, operating with
a given toroid and solenoid polarity, introduces
an apparent dimuon charge asymmetry of approximately 0.006
in absolute value
(to be discussed later in this Section).
This detector effect
changes sign when the toroid and
solenoid polarities are reversed (since the exact same
track that is called \textquotedblleft{positive}"
with one polarity 
is called \textquotedblleft{negative}"
with the opposite polarity).
This effect can be seen by comparing the $\eta$ histograms
in Figs. \ref{pt_eta_phi_opp} and \ref{pt_eta_phi_same}.
Therefore, to cancel detector geometry effects to first order, 
we always consider
data sets that have equal event counts
for each toroid-solenoid magnet polarity (or weight the events appropriately).
We combine events with one solenoid polarity and
toroid polarity, with events with the opposite solenoid polarity and
toroid polarity. The analysis is done separately for solenoid polarity
equal to the toroid polarity, and solenoid polarity
opposite to the toroid polarity.

Let $n_\alpha^{\beta\gamma}$ be the number of muons passing cuts with charge
$\alpha = \pm 1$, toroid polarity $\beta = \pm 1$, and
$\gamma = +1$ if
$\eta > 0$ and $\gamma = -1$ if $\eta < 0$. 

We model the physics and
the detector as follows:
{\small
\begin{eqnarray}
n_\alpha^{\beta\gamma} & \equiv & \frac{1}{4}
N \epsilon^\beta (1 + \alpha A)(1 + \alpha \gamma A_{fb})
(1 + \gamma A_{det}) \nonumber \\
& & \times 
(1 + \alpha\beta\gamma A_{ro})
(1 + \beta \gamma A_{\beta \gamma})
(1 + \alpha \beta A_{\alpha \beta}),
\label{asymmetries}
\end{eqnarray}}
where $\epsilon^+ + \epsilon^- = 1$.
The eight equations (\ref{asymmetries})
\textit{define} eight parameters in terms of the 
eight numbers $n_\alpha^{\beta\gamma}$.
The parameters are $N$, $\epsilon^+$, and
six asymmetries.
$N \epsilon^\beta$ is approximately equal to the number of
muons passing cuts with toroid polarity $\beta$.
$A$ is the dimuon charge asymmetry, 
$A_{fb}$ is the forward-backward
asymmetry (that quantifies the tendency of $\mu^+$ to go in
the proton direction and $\mu^-$ to go in the anti-proton
direction), $A_{det}$ 
measures the north-south asymmetry
of the detector (\textquotedblleft{north}" has $\eta < 0$), and $A_{ro}$ is the range-out asymmetry
(that quantifies the change in acceptance and range-out of
muon tracks that bend toward, or away from, the beam line,
see Fig. \ref{detector}).
$A_{\alpha \beta}$ is a detector asymmetry between
tracks bending north and tracks bending south.
$A$ and $A_{fb}$ are physics
asymmetries that we want to measure, and $A_{det}$,
$A_{ro}$ and $A_{\alpha \beta}$ are detector asymmetries.
$A_{\beta \gamma}$ is a second-order asymmetry that is different
from zero only if $A$ \textit{and} $A_{ro}$
are different from zero.
If the selection of events includes single 
muon \textit{and} dimuon cuts,
we use capital $A$ for the asymmetries.
If only single muon cuts are required,
we use lower case $a$. 
In Tables \ref{dimu-} and \ref{dimu+} we show the numbers 
$n_\alpha^{\beta \gamma}$ for our standard cuts.
The measured asymmetries are presented in Table \ref{asym7}.

We can understand the detector asymmetry $A_{\alpha \beta}$
in more detail. Let $A^\beta$ be the dimuon charge
asymmetry of events with toroid polarity $\beta$.
From Eq.~(\ref{asymmetries}) we obtain, to first order
in the asymmetries, $A^\beta = A + \beta A_{\alpha \beta}$.
Therefore $\beta A_{\alpha \beta}$ is the dimuon charge
asymmetry due to detector geometry effects. It changes sign
when the magnet polarities are reversed. This is the detector
geometry effect that is cancelled by taking the weighted
average of $A^+$ and $A^-$.
The magnitude of this effect is 
$A_{\alpha \beta} \approx -0.006$, see Table \ref{asym7}.

Solving equations (\ref{asymmetries}) up to
second-order terms in the asymmetries,
 $A$ is obtained by taking
the weighted average $\frac{1}{2} (A^+ + e A^-)$:
{\small
\begin{eqnarray}
\frac{  (n_+^{++} + n_+^{+-} - n_-^{++} - n_-^{+-})
    + e (n_+^{-+} + n_+^{--} - n_-^{-+} - n_-^{--})}
     {  (n_+^{++} + n_+^{+-} + n_-^{++} + n_-^{+-})
    + e (n_+^{-+} + n_+^{--} + n_-^{-+} + n_-^{--})} \nonumber \\
\qquad = A + A_{fb} A_{det},
\label{A}
\end{eqnarray}
}
\noindent
where $e \equiv \epsilon^+ / \epsilon^-$.
To the required accuracy, $e$ is the ratio of the
number of events passing cuts with toroid polarity $\beta = +1$
over the corresponding number with $\beta = -1$.
The ratio $e$ is determined by
counting single muons for the single muon asymmetries,
or by counting dimuons for the dimuon analysis.
This procedure introduces no bias since we count all 
muons or dimuons regardless
of the charges of each muon.
The left-hand-side is the measured asymmetry, $A$ is the
corrected asymmetry, and $-A_{fb} A_{det}$ is the correction
due to the forward-backward \textit{and} detector asymmetries.
We do not apply this correction because it turns out to
be negligible and compatible with zero, 
see Table \ref{asym7}. We use the correction to
estimate the corresponding systematic uncertainty.
We can understand the last term in Eq.~(\ref{A}): if
positive (negative) muons prefer to go in the proton
(antiproton) direction \textit{and} the detector is north-south
asymmetric, then we obtain an apparent charge asymmetry.

The forward-backward asymmetry $A_{fb}$ (up to second-order
terms) is:
{\small
\begin{eqnarray}
\frac{  (n_+^{++} + n_-^{+-} - n_+^{+-} - n_-^{++})
    + e (n_+^{-+} + n_-^{--} - n_+^{--} - n_-^{-+})}
     {   (n_+^{++} + n_-^{+-} + n_+^{+-} + n_-^{++})
    + e (n_+^{-+} + n_-^{--} + n_+^{--} + n_-^{-+})} \nonumber \\
= A_{fb} + A A_{det}.
\label{Afb}
\end{eqnarray}
}

\begin{table}
\caption{Numbers $n_\alpha^{\beta\gamma}$ of muons
passing standard single and dimuon
cuts with charge $\alpha$, toroid
magnet polarity $\beta$, and  $\eta< 0$ ($\gamma = -1$)
or $> 0$ ($\gamma = +1$). There are two entries per event.
The solenoid and toroid polarities are opposite.
In total there are 154667 positive-positive, 154482 negative-negative, and
1075192 positive-negative dimuon events. 
No cuts on the dimuon mass are imposed at this stage.}
\begin{ruledtabular}
\begin{tabular}{cccc}
charge & toroid & $-2.2 < \eta < 0.0$ & $0.0 < \eta < 2.2$ \\
$\alpha$ & polarity $\beta$ & $\gamma = -1$ & $\gamma = +1$ \\
\hline
$+1$ & $+1$ & $367376$ & $335700$ \\
$-1$ & $+1$ & $348295$ & $353453$ \\
$+1$ & $-1$ & $337697$ & $343753$ \\
$-1$ & $-1$ & $356891$ & $325517$ \\
\end{tabular}
\end{ruledtabular}
\label{dimu-}
\end{table}

We have repeated the study of detector asymmetries for
the central ($\left| \eta \right| < 0.95$) and
forward ($0.95 < \left| \eta \right| < 2.2$) muon systems
separately.

Also shown in Table \ref{asym7} is the dimuon charge asymmetry
for flavor creation $A_{fc}$ 
(defined as $\Delta \psi \ge 90^\circ$)
and flavor excitation $A_{fe}$
(defined as $\Delta \psi < 90^\circ$),
where $\Delta \psi$ is the 3-dimensional angle between the two 
muons.
Flavor creation corresponds
to the $b$ and $\bar{b}$ quarks in opposite jets, while 
flavor excitation corresponds generally to a
$b \bar{b}$ pair produced in the hadronization of one parton,
including gluon splitting.
The accepted cross sections for flavor creation
and flavor excitation are nearly equal
(as indicated by their statistical errors in Table \ref{asym7}).
In Table \ref{asym7}  
we also show the ratio $R = (N^{++} + N^{--})/N^{+-}$
of like-sign to opposite-sign dimuon events in either the mass window
5.0 to 8.7 GeV/$c^2$ or 11.5 to 30 GeV/$c^2$
for events passing a dimuon trigger. This ratio is used for mixing
studies. These mass windows reduce backgrounds from same-side 
direct-sequential muon pairs (process $P_4$
in Table \ref{P}), and decays of $J/\psi$'s, $\Upsilon$'s 
and their resonances (part of process $P_6$), defined later. 

\begin{table}
\caption{Same as Table \ref{dimu-}, but for
equal solenoid and toroid polarities.
In total there are 136422 positive-positive, 
136857 negative-negative, and
944013 positive-negative dimuon events.}
\begin{ruledtabular}
\begin{tabular}{cccc}
charge & toroid & $-2.2 < \eta < 0.0$ & $0.0 < \eta < 2.2$ \\
$\alpha$ & polarity $\beta$ & $\gamma = -1$ & $\gamma = +1$ \\
\hline
$+1$ & $+1$ & $306594$ & $279508$ \\
$-1$ & $+1$ & $290270$ & $296264$ \\
$+1$ & $-1$ & $311153$ & $319602$ \\
$-1$ & $-1$ & $329285$ & $301908$ \\
\end{tabular}
\end{ruledtabular}
\label{dimu+}
\end{table}

\begin{table}
\caption{
Asymmetries 
described in Section IV 
are shown for central (c) muons, $|\eta| < 0.95$, and forward (f)
muons, $0.95 < |\eta| < 2.2$.
All errors are statistical. 
The average for opposite and equal toroid and solenoid polarities
(\textsf{torpol$\cdot$solpol} = $-1$ and 1) is $A = -0.0005 \pm 0.0013$ (all).
}
\begin{ruledtabular}
\begin{tabular}{lrr}
\textsf{torpol*solpol} & $-1$ & $1$ \\
\hline
$e$               & $1.0300$  & $0.9292$ \\
\hline
$a$ (all)         & $0.0001 \pm 0.0005$  &   $-0.0012 \pm 0.0005$ \\
$a_{fb}$ (all)    & $0.0021 \pm 0.0005$  &   $0.0011 \pm 0.0005$ \\
$a_{det}$ (all)   & $-0.0074 \pm 0.0005$  &   $-0.0045 \pm 0.0005$ \\
$a_{ro}$ (all)    & $-0.0268 \pm 0.0005$  &   $-0.0298 \pm 0.0005$ \\
$a$ (c)           & $-0.0014 \pm 0.0007$  &   $-0.0035 \pm 0.0008$ \\
$a_{fb}$ (c)      & $0.0004 \pm 0.0007$  &   $-0.0010 \pm 0.0008$ \\
$a_{det}$ (c)     & $-0.0069 \pm 0.0007$  &   $-0.0068 \pm 0.0008$ \\
$a_{ro}$ (c)      & $-0.0867 \pm 0.0007$  &   $-0.0891 \pm 0.0008$ \\
$a$ (f)           & $0.0010 \pm 0.0006$  &   $0.0003 \pm 0.0006$ \\
$a_{fb}$ (f)      & $0.0033 \pm 0.0006$  &  $0.0025 \pm 0.0006$ \\
$a_{det}$ (f)     & $-0.0078 \pm 0.0006$  &  $-0.0029 \pm 0.0006$ \\
$a_{ro}$ (f)      & $0.0122 \pm 0.0006$  &  $0.0089 \pm 0.0006$ \\
\hline
$A_{fb}$ (all)    & $0.0006 \pm 0.0006$  &   $0.0001 \pm 0.0007$ \\
$A_{det}$ (all)   & $-0.0187 \pm 0.0006$  &   $-0.0165 \pm 0.0007$ \\
$A_{ro}$ (all)    & $-0.0268 \pm 0.0006$  &   $-0.0283 \pm 0.0007$ \\
$A_{\alpha \beta}$ (all)& $-0.0059 \pm 0.0006$ &  $-0.0069 \pm 0.0007$ \\
$A_{\beta \gamma}$ (all)& $-0.0002 \pm 0.0006$ &  $-0.0015 \pm 0.0007$ \\
\hline
$R$         & $0.4603 \pm 0.0013$   &  $0.4610 \pm 0.0013$ \\
\hline
$A_{fc}$ (all)    & $-0.0008 \pm 0.0026$   &   $-0.0023 \pm 0.0027$ \\
$A_{fe}$ (all)    & $0.0019 \pm 0.0026$   &   $-0.0008 \pm 0.0028$ \\
$A$ (cc)          & $0.0015 \pm 0.0044$  &   $-0.0021 \pm 0.0048$ \\
$A$ (ff)          & $0.0024 \pm 0.0031$  &   $0.0016 \pm 0.0033$ \\
$A$ (all)         & $0.0005 \pm 0.0018$  &   $-0.0016 \pm 0.0019$ \\
\end{tabular}
\end{ruledtabular}
\label{asym7}
\end{table}

\section{Dimuon processes}
In this Section we obtain the factor $f$ defined in Eq.~(\ref{epsilon}).
We consider the processes $P_1$ -- $P_{13}$
listed in Table \ref{P}.
$P_1$ is direct-direct $b \bar{b}$ decay.
$P_2$ is opposite-side direct-sequential decay.
$P_3$ is sequential-sequential decay.
$P_4$ is same-side direct-sequential decay.
$P_7$ corresponds to cosmic ray muons that
traverse the D0 detector and are reconstructed
twice: once upon entry and once upon exit.
$P_8$ corresponds to muons from $K^{\pm}$
decay, in coincidence with a prompt muon from
the collision. This process is discussed in
detail in Sections VI and VII.
$P_9$ corresponds to a cosmic ray muon, in
coincidence with a prompt muon.
$P_{10}$ corresponds to a hadron that traverses
the calorimeter and iron toroid and is reconstructed
as a muon, in coincidence with a prompt muon.
$P_{11}$ corresponds to a combinatoric background
faking a muon, in coincidence with a prompt muon.
Examples of \textquotedblleft{other}" processes $P_{12}$ are
dimuons with the following parents:
$B^\pm$ and $\pi^\pm$, $\bar{B}^0$ and $\tau^\pm$, $B_s^0$ and $J/\psi$,
$B^0$ and $\tau^\pm$, $\bar{B}^0$ and $J/\psi$, $B^\pm$ and $\tau^\pm$,
$b$ and unrelated $c$. 
$P_{13}$ are events that have one of the muons
reconstructed with the wrong sign.
Contributions from both muons coming from hadron misidentification
or combinatoric background are negligible.

Let $\chi_d$ be the probability that a $\bar{B}^0(b \bar{d})$ 
meson that decays to a flavor specific final state,
mixes and decays
as a $B^0(d \bar{b})$. Similarly, $\bar{\chi}_d$ is the probability
that a $B^0$ meson mixes and decays as a $\bar{B}^0$.
Here we consider the possibility that
$\chi_d \ne \bar{\chi}_d$ (the general case with 
$\chi_d \ne \bar{\chi}_d$ and $\chi_s \ne \bar{\chi}_s$ is considered
in Section IX).
The probability that a $b$ quark in the sample
decays as a $\bar{b}$ is
\begin{equation}
\chi = f_d \frac{\beta_d}{\left< \beta \right>} \chi_d
     + f_s \frac{\beta_s}{\left< \beta \right>} \chi_s,
\label{chi}
\end{equation}
where $f_d$ and $f_s$ are the fractions of $b$ quarks that hadronize
to $B^0$ or $\bar{B}^0$ and $B_s^0$
or $\bar{B}_s^0$ respectively, and $\beta_d$, $\beta_s$
and $\left< \beta \right>$ are the branching fractions for
$B^0$, $B_s^0$ and the $b$-hadron admixture respectively
decaying to $\mu X$ with $\mu$ passing cuts. Similarly,
\begin{equation}
\bar{\chi} = f_d \frac{\beta_d}{\left< \beta \right>} \bar{\chi}_d
     + f_s \frac{\beta_s}{\left< \beta \right>} \chi_s
\label{chibar}
\end{equation}
is the probability that a $\bar{b}$ in the sample
decays as a $b$.
From Ref.~\cite{pdg} we take $f_d = 0.397 \pm 0.010$,
$f_s = 0.107 \pm 0.011$, 
$\chi_{d0} \equiv \frac{1}{2} (\chi_d + \bar{\chi}_d) = 0.186 \pm 0.004$,
and $\chi_{s0} \equiv \frac{1}{2} (\chi_s + \bar{\chi}_s) > 0.49883$. 
We take $\beta_d = \beta_s = \left< \beta \right>$.
To abbreviate, we define
$\chi_0 \equiv \frac{1}{2} \left( \chi + \bar{\chi} \right)$,
and $\xi \equiv 2 \chi_0 \left( 1 - \chi_0 \right)$.
We assume CPT symmetry.
The probability that a $B$ ($\bar{B}$) hadron that
decays to a flavor specific final state, decays as a $B$ ($\bar{B}$)
is then $1 - \chi_0$.
From Table \ref{P} we obtain 
the dimuon charge asymmetry $A$ after correcting for
asymmetric kaon decay, (i.e.,  after subtracting
a term $0.5 a P_8$ in the numerator),
\begin{equation}
A = \frac{(\chi - \bar{\chi}) \left[ (1 - \chi_0) (P_1 - P_3) + 0.25 \rho' P'_8 \right]}
{A_{den}},
\label{Afull}
\end{equation}
and the factor $f$ in Eq.~(\ref{epsilon}):
\begin{equation}
f = \frac{A_{den}}
{8 f_d \chi_{d0} \left[ (1 - \chi_0) (P_1 - P_3) + 0.25 \rho' P'_8 \right]},
\label{f}
\end{equation}
where 
$A_{den} = \xi (P_1 + P_3) + (1 - \xi) P_2 + 
0.28 P_7 + 0.5 P'_8 + 0.78 P_{13}$
and $P'_8 \equiv P_8 + P_9 + P_{10} + P_{11}+ P_{12}$.
The fraction of prompt
muons from $b$ decay is $\rho' \equiv 0.6 \pm 0.15$ \cite{muon2}.

\section{Weights $P_i$ of dimuon processes}

The weights $P_2$ through $P_{13}$, normalized to
direct-direct $b \bar{b}$ decay $P_1 \equiv 1$,
are summarized in Table \ref{dimu_Ps}.
The weights $P_2$, $P_4$, $P_5$, $P_6$ and $P_{12}$ 
were obtained from
Monte Carlo simulations
(using the {\sc{Pythia}} generator \cite{pythia})
with full detector
simulation (based on the {\sc{Geant}} program \cite{D0Geant}) 
and event reconstruction and selection.
A cross-check for weight $P_2$ is the measurement of
the average mixing probability of $B$ hadrons to be 
described below. Weights $P_4$, $P_5$ and $P_6$
do not contribute like-sign dimuons, and so  do not
enter into the measurement of the CP-violation parameter.
Weight $P_3$ was obtained from $P_3 \approx P_2^2/(4 P_1)$.
Weight $P_7$ was obtained by two methods:
(i) from the data of a cosmic ray run, and
(ii) extrapolating the out-of-time muon background
(as measured by the scintillators) into the
acceptance window of $\pm5$~ns.
Weight $P_9$ was obtained using the data of the
cosmic ray run.
Weight $P_{10}$ was obtained by counting the number
of tracks that had enough momentum to traverse the
calorimeter and iron toroid, and multiplying by the
probability $\exp{(-14)}$ that they do not interact
(the calorimeter has $\approx 7$ nuclear interaction
lengths, and the iron toroid 
has $\approx 7$ nuclear interaction lengths).
Weight $P_{11}$ was estimated by relaxing the number of
required wire chamber hits (one less hit in layer \textsf{A}
and/or 1 less hit in layers \textsf{B} or \textsf{C}). 
Weight $P_{13}$ was estimated using the measured
resolution of the local muon spectrometer.
In our data set, passing standard single and dimuon 
cuts, we expect $\approx 1$ dimuon
event from $Z$ boson decay, and less than one event
from prompt muons plus $W^\pm$ boson decay.

Let us consider the weight $P_8$ in some detail.
This weight corresponds to prompt muons from $b$ 
or $c$ or $s$ decay plus $K^\pm$ decay.
This is an important background because kaon 
interactions are charge asymmetric, and dominate
the systematic uncertainty of the measurement of the
CP-violation parameter.
The inelastic interaction length of $K^+$ in the calorimeter
is greater than the inelastic interaction length of $K^-$.
This difference is due to the
existence of hyperons $Y$ (strangeness $-1$ baryons: $\Lambda$,
$\Sigma$, $Y^*$). Reactions $K^- N \rightarrow Y \pi$ have
no $K^+ N$ analog.
Therefore $K^+$ has more time to decay than $K^-$.
The result is a charge asymmetry from $K^\pm$ decay.
The single muon charge asymmetry from $K^\pm$ decay
is obtained from the inelastic cross sections for $K^- d$ 
and $K^+ d$ \cite{pdg} and the geometry and materials
of the D0 detector:
$a \equiv (n^+ - n^-)/(n^+ + n^-) = 0.026 \pm 0.005$.

\begin{table*}
\caption{Processes contributing to dimuon events.
Each row includes processes related by CP conjugation and $B \leftrightarrow \bar{B}$
mixing. The weights are normalized to direct-direct $b \bar{b}$ decay
$P_1 \equiv 1$.
$\xi \equiv 2 \chi_0 \left( 1 - \chi_0 \right)$.
$\chi = f_d \chi_d + f_s \chi_s$ is the probability that $b$ quarks
decay as $\bar{b}$.
$\bar{\chi} = f_d \bar{\chi}_d + f_s \chi_s$ is the probability that $\bar{b}$
anti-quarks decay as $b$. The fraction of prompt
muons from $b$ decay is $\rho' \equiv 0.6 \pm 0.15$~\cite{muon2}.
$\rho \equiv \frac{1}{2} \rho'(\chi - \bar{\chi})$. 
$a = 0.026 \pm 0.005$
is the charge asymmetry of $K^\pm$ decay, see the text.
CPT symmetry is assumed. 
For example, the number of direct-direct decays 
$b \bar{b} \rightarrow \mu^+ \mu^+ X$ is $\propto P_1 \chi (1 - \chi_0)$.}
\begin{ruledtabular}
\begin{tabular}{ccccc}
process & weight & $N^{++}$ & $N^{--}$ & $N^{+-}$ \\
\hline
$b \rightarrow \mu^-$, $\bar{b} \rightarrow \mu^+$ &
$P_1 \equiv 1$ & $\chi (1 - \chi_0)$
& $(1 - \chi_0) \bar{\chi}$ & $1 - \xi$ \\
$b \rightarrow \mu^-$, $\bar{b} \rightarrow \bar{c} \rightarrow \mu^-$ & $P_2$
& $\frac{1}{2}(1 - \xi)$ & $\frac{1}{2}(1 - \xi)$
& $\xi$ \\
$b \rightarrow c \rightarrow \mu^+$, $\bar{b} \rightarrow \bar{c} \rightarrow \mu^-$
& $P_3$ & $(1 - \chi_0) \bar{\chi}$ & $\chi (1 - \chi_0)$
& $1 - \xi$ \\
$b \rightarrow \mu^- c \rightarrow \mu^+$ & $P_4$ & 0 & 0 & 1 \\
$c \rightarrow \mu^+$, $\bar{c} \rightarrow \mu^-$ & $P_5$ & 0 & 0 & 1 \\
Drell-Yan, $J/\psi$, $\Upsilon$ & $P_6$ & 0 & 0 & 1 \\
dimuon cosmic rays & $P_7$ & $\approx 0.14$ & $\approx 0.14$ & $\approx 0.72$ \\
$\mu$ + $K^\pm$ decay & $P_8$ & $0.25 \cdot (1+a+\rho)$ & $0.25  \cdot (1-a-\rho)$ & $0.5$ \\
$\mu$ + cosmic & $P_9$ & $\approx 0.25 \cdot (1+\rho)$ & $\approx 0.25 \cdot (1-\rho)$ & $\approx 0.5$ \\
$\mu$ + punch-through & $P_{10}$ &  $0.25 \cdot (1+\rho)$ & $0.25 \cdot (1-\rho)$ & $0.5$ \\
$\mu$ + combinatoric & $P_{11}$ & $0.25 \cdot (1+\rho)$ & $0.25 \cdot (1-\rho)$ & $0.5$ \\
other &  $P_{12}$ & $0.25 \cdot (1+\rho)$ & $0.25 \cdot (1-\rho)$ & $0.5$ \\
dimuon w. wrong sign & $P_{13}$ & 0.39 & 0.39 & 0.22 \\
\end{tabular}
\end{ruledtabular}
\label{P}
\end{table*}

\begin{table}
\caption{Weights of dimuon processes for standard cuts
(obtained as described in the text). 
Note that 64\% of dimuons
are from direct-direct $b \bar{b}$ decay.}
\begin{ruledtabular}
\begin{tabular}{cc}
$P_1$ & $\equiv 1$ \\
$P_2$ & $0.116 \pm 0.055$ \\
$P_3$ & $0.003 \pm 0.003$ \\
$P_4$ & $0.093 \pm 0.049$ \\
$P_5$ & $0.070 \pm 0.042$ \\
$P_6$ & $0.023 \pm 0.023$ \\
$P_7$ & $0.003 \pm 0.003$ \\
$P_8$ & $0.078 \pm 0.023$ \\
$P_9$ & $ 0.0001 \pm 0.0001$ \\
$P_{10}$ & $0.001 \pm 0.001$ \\
$P_{11}$ & $0.0002 \pm  0.0002$ \\
$P_{12}$ & $0.163 \pm 0.066$ \\ 
$P_{13}$ & $0.0005 \pm 0.0005$ \\
\end{tabular}
\end{ruledtabular}
\label{dimu_Ps}
\end{table}

For $P_8$, we make two complementary estimates
based on data. In the first, we measure
the exclusive decay
$B^0 \rightarrow D^*(2010)^- \mu^+ \nu_\mu$,
$D^*(2010)^- \rightarrow \bar{D}^0 \pi^-$,
$\bar{D}^0 \rightarrow K^+ \pi^-$, and
its charge conjugate. 
We apply standard single and dimuon cuts, and
count events with and without a muon matching the kaon track.
We subtract the background by two methods: using a side-band
of $m_{D^{*-}} - m_{D^0}$, or using the
wrong relative sign of the muon from $B^0$ decay and
the pion from $D^{*-}$ decay.
The result is $P_8 = 0.078 \pm 0.013\textrm { (stat)} \pm 0.019\textrm { (syst)}$.
From studies with this exclusive decay,
we learn that the 
global track $\chi^2$-cut is not very effective in
reducing $K^\pm$ decay kinks (for the high momentum muons
passing cuts). Therefore we must correct $A$
for $K^\pm$ decay as discussed in Section VII.

We use the following alternative procedure to estimate 
the background weight $P_8$ from data. Instead of
$K^+ \rightarrow \mu^+ \nu$, we study
$K^0_S \rightarrow \pi^+ \pi^-$. 
For this estimate,
we assume that the production and decay kinematics 
of $K^+$ and $K^0_S$ are
approximately the same. The branching fractions are similar.
To account for the smaller decay length of $K^0_S$ compared to $K^\pm$,
we scale the volume in which
the $K^\pm$ can decay (719 mm in radius and 1360 mm in half-length)
by the fraction of lifetimes.
$719$ mm is the sum of the inner radius of the calorimeter, plus
the transverse interaction length of $K^+$ in the
calorimeter.
We analyze single muon (instead of dimuon) data to account for
correlations.
We compare two histograms: one is
$p_T$ of pions from
$K^0_S$'s decaying in the scaled-down volume,
and the other histogram is $p_T$ of the second 
(in order of decreasing $p_T$) muon passing cuts.
By this indirect method we obtain
$P_8 \approx 0.041 \pm 0.010\textrm{(stat)} \pm 0.041\textrm{(syst)}$.

Within errors, the two measurements of $P_8$
agree and we use the result from the first method.

The systematic uncertainty of $P_8$, $\pm 0.019$, was
taken as half the difference of 0.078 and 0.041.
This value is reasonable in view of the variation
of the measured $P_8$ with different cuts and
data sub-sets.

From Monte Carlo simulations we obtain
$P_8 = 0.047 \pm 0.034\textrm{ (stat)}$, consistent
with the above. 

\section{Systematic uncertainties of $A$}
We add in quadrature the following systematic uncertainties.
A summary is presented in Table \ref{DA}.

\textit{Detector effects.}
Before averaging over magnetic field polarities, 
the detector introduces a dimuon charge 
asymmetry of  $\approx0.006$ in absolute value,
as discussed in Section IV. 
After averaging over magnetic field polarities 
(with appropriate weights), the uncertainty of the dimuon charge
asymmetry due to detector geometry effects is
$|A_{fb} \cdot A_{det}|$ (see Eq.~(\ref{A})).
As a measure of this uncertainty, we have used
the largest deviation from zero,
$|A_{fb} \cdot A_{det}| = 0.0049 
\cdot 0.030 = 0.00015$, obtained after any of the 54 sets of
dimuon cuts, for any of the three detector regions
(central, forward, or all).

\textit{Inaccuracy of $e \equiv \epsilon^+ / \epsilon^-$.}
We have obtained the ratio $e$ by counting dimuon events with
toroid polarity $\beta = 1$ and dividing by the corresponding
number for $\beta = -1$. 
This procedure introduces no bias since we count all dimuons regardless
of the charges of each muon.
We take $\Delta e = 0.03$ from the largest difference between any cuts.
Multiplying by the detector dimuon charge asymmetry 
before averaging over magnet polarities, $\approx 0.006$
in absolute value, 
we obtain $\Delta A \approx 0.00018$.

\textit{Prompt $\mu$ + $K^\pm$ decay.}
The single muon charge asymmetry of kaon decay is $a = 0.026 \pm 0.005$
as explained in Section VI.
We take $P_8 = 0.078 \pm 0.023$ from Table \ref{dimu_Ps}.
The corresponding correction to $A$, explained in Section V, is
$\delta A = -0.5 a P_8 / A_{den} =
-0.5 \times (0.026 \pm 0.005) \times
(0.078 \pm 0.023) / A_{den} = -0.0023 \pm 0.0008$
($A_{den} \approx 0.436$ 
is the denominator of Eq.~(\ref{Afull})).
This uncertainty on $\delta A$ is by far the dominating 
systematic uncertainty of the entire measurement.

\textit{Dimuon cosmic rays.}
These are cosmic rays detected twice, once as they enter and
once as they exit the D0 detector.
We take $P_7 < 0.007$ from Table \ref{dimu_Ps}.
From cuts that select cosmic rays, we obtain the 
apparent dimuon charge
asymmetry $A = -0.0095 \pm 0.0117$.
Then the corresponding uncertainty of $A$ is
$< 0.007 \times 0.28 \times 0.021 / A_{den} = 0.0001$
(see Table \ref{P}; $0.021 = |-0.0095| + 0.0117$).

\begin{table}
\caption{Systematic uncertainties of the dimuon charge asymmetry $A$
for standard cuts.}
\begin{ruledtabular}
\begin{tabular}{cc}
Source of error & $\Delta A$ \\
\hline
detector                      & 0.00015 \\
$e = \epsilon^+ / \epsilon^-$ & 0.00018 \\
prompt $\mu$ + $K^\pm$ decay  & 0.00083 \\
dimuon cosmic rays            & 0.00010 \\
prompt $\mu$ + cosmic $\mu$   & 0.00001 \\
wrong charge sign             & 0.00018 \\
punch-through                 & 0.00001 \\
\hline
Total                         & 0.00089 \\
\end{tabular}
\end{ruledtabular}
\label{DA}
\end{table}

\textit{Prompt $\mu$ + single cosmic ray.} 
We take $P_9 < 0.0002$,
see Table \ref{dimu_Ps}. From cuts that select cosmic rays,
we obtain an apparent 
single muon charge asymmetry $a = 0.026 \pm 0.002$.
Then the uncertainty in $A$ is 
$< 0.0002 \times 0.5 \times 0.028 / A_{den} = 6 \times 10^{-6}$
(see Table \ref{P}; $0.028 = 0.026 + 0.002$).

\textit{Wrong local muon sign.}
$P_{13} < 0.001$, see Table \ref{dimu_Ps}. 
Even if the asymmetry of wrong tracks is 0.1
(overestimate), the corresponding uncertainty of $A$ is small:
$< 0.001 \times 0.78 \times 0.1 / A_{den} = 0.00018$
(see Table \ref{P}).

\textit{Punch-through.}
We take $P_{10} < 0.002$, see Table \ref{dimu_Ps}. 
The measured charge asymmetry of tracks with $p_T > 3.0$ GeV/$c$ is
$0.0049 \pm 0.0005$ due to showers on matter instead of antimatter.
Then the error in $A$ is
$< 0.002 \times 0.5 \times 0.0054 / A_{den} = 1 \times 10^{-5}$
(see Table \ref{P}; $0.0054 = 0.0049 + 0.0005$).

\begin{table}
\caption{Systematic uncertainties of $f$.}
\begin{ruledtabular}
\begin{tabular}{cl}
Source of error & $\Delta f$ \\
\hline
$P_2$ & 0.084 \\
$P_7$ & 0.002 \\
$P'_8$ & 0.056 \\
$\chi_{d0}$ & 0.012 \\
$f_d$ & 0.014 \\
$\chi_{s0}$ & 0.0002 \\
$f_s$ & 0.019 \\
$\rho'$ & 0.008 \\
$P_{13}$ & 0.0007 \\
\hline
Total  & 0.105 \\
\end{tabular}
\end{ruledtabular}
\label{errors11}
\end{table}

\section{Other cross-checks}
The sign of central muons was cross-checked using cosmic rays
(which are charge asymmetric \cite{pdg}).
The sign of forward muons relative to the sign of central muons
was cross-checked with $J/\psi$'s.

The direct and reverse magnetic fields in the iron toroid
were measured to be equal to within 0.1\%.

We find the dimuon charge asymmetry $A$ stable (within statistical errors)
for all recorded events or events passing a set of dimuon triggers, and
across the different cuts (54 sets were studied), 
data subsets, opposite or equal toroid and
solenoid polarities, central or forward muons, or flavor creation or
flavor excitation events.

\section{Results}
We obtain
$A = 0.0005 \pm 0.0018\textrm{ (stat)}$
for opposite solenoid and toroid polarities, and
$A = -0.0016 \pm 0.0019\textrm{ (stat)}$
for equal solenoid and toroid polarities (see last line in 
Table \ref{asym7}).
Combining these measurements we obtain
\begin{equation}
A = -0.0005 \pm 0.0013\textrm{ (stat)}.
\label{final_A_4}
\end{equation}
We add a correction
$\delta A = -0.0023 \pm 0.0008$
to $A$ due to asymmetric $K^\pm$ decay
(this effect is explained in Sections VI and VII). 
The uncertainty of this correction dominates the 
systematic uncertainties of the CP-violation parameter.
The final corrected value of the
dimuon charge asymmetry is
\begin{equation}
A = -0.0028 \pm 0.0013\textrm{ (stat)} \pm 0.0009\textrm{ (syst)}.
\label{final_A_5}
\end{equation}
The breakdown of systematic uncertainties of $A$ is presented in Table \ref{DA}.

From the dimuon charge asymmetry $A$ we obtain
{
\begin{eqnarray}
\frac{\Re{(\epsilon_{B^0})}}{1 + \left| \epsilon_{B^0} \right|^2} 
& = & \frac{A_{B^0}}{4} \equiv f \cdot A \\
& = &  -0.0023 \pm 0.0011\textrm{ (stat)} \pm 0.0008\textrm{ (syst)}, \nonumber
\label{CPV_final}
\end{eqnarray}
}
where
\begin{equation}
f = 0.814 \pm 0.105\textrm{ (syst)}.
\label{f0}
\end{equation}
The breakdown of systematic uncertainties of $f$,
calculated from information provided
in preceding sections, is listed in
Table \ref{errors11}.
In comparison, the Particle Data Group average of 
2004 \cite{pdg} is 
${\Re{(\epsilon_{B^0})}}/{(1 + \left| \epsilon_{B^0} \right|^2)} =
0.0005 \pm 0.0031$.

All preceding equations correspond to the 
case $\chi_s = \bar{\chi}_s$.
From (\ref{Afull}) we obtain, for the general case
$\chi_d \ne \bar{\chi}_d$ and $\chi_s \ne \bar{\chi}_s$,
\begin{equation}
A = \frac{1}{4 f} \left[ A_{B^0}
+ \frac{f_s \chi_{s0}}{f_d \chi_{d0}} A_{B^0_s} \right].
\label{Ageneral}
\end{equation}

We measure the ratio $R$ of like-sign to opposite-sign
dimuons. For this measurement we
require the invariant mass of the two muons to be either in the
window 5.0 to 8.7 GeV/$c^2$ or 11.5 to 30 GeV/$c^2$.
These mass cuts are designed to reduce backgrounds from
same-side direct-sequential decay and backgrounds from
$\psi$ and $\Upsilon$ meson decays to allow a measurement of
$B \leftrightarrow \bar{B}$ mixing. 
For this reason we set
$P_4 = P_6 = 0$ for the mixing analysis. 
We also require dimuon triggers
from a list that excludes triggers requiring opposite sign
muons.
We obtain
$R = 0.461 \pm 0.001\textrm{ (stat)} \pm 0.010\textrm{ (syst)}$, and
$\xi = 0.229 \pm 0.001\textrm{ (stat)} \pm 0.036\textrm{ (syst)}$.
The breakdown of systematic uncertainties,
calculated from information provided
in preceding sections,
is shown in Table \ref{errors_mixing}.
The final result for the mixing probability, averaged over the mix of
hadrons with a $b$ quark, is
\begin{equation}
\chi_0 = 0.132 \pm 0.001\textrm{ (stat)} \pm 0.024\textrm{ (syst)}.
\label{chi_1}
\end{equation}
The agreement with the world average, $0.127 \pm 0.006$ \cite{pdg},
is a sensitive test of $P_2$ and $f$, since the largest
systematic uncertainty of $\xi$ and $f$ are due to the same
weight $P_2$.

\begin{table}
\caption{Systematic uncertainties of 
$\xi = 2 \chi_0 ( 1 - \chi_0)$.
$P'_4 \equiv P_4 + P_5 + P_6$.}
\begin{ruledtabular}
\begin{tabular}{cl}
Source of error & $\Delta \xi$ \\
\hline
$R$ & 0.0074 \\
$P_2$ & 0.028 \\
$P'_4$ & 0.015 \\
$P_7$ & 0.0001 \\
$P'_8$ & 0.015 \\
$P_{13}$ & 0.0002 \\
\hline
Total  & 0.036 \\
\end{tabular}
\end{ruledtabular}
\label{errors_mixing}
\end{table}

Finally, we measure the tendency of $\mu^+$ ($\mu^-$) to go in
the proton (antiproton) direction. We obtain the forward-backward
asymmetry for events passing standard single and dimuon cuts:
\begin{equation}
A_{fb} = 0.0004 \pm 0.0005\textrm{ (stat)} \pm 0.0002\textrm{ (syst)}.
\label{Afb_dimuon}
\end{equation}
$A_{fb}$ is defined in Section IV.
The systematic uncertainty 
is $|A| |A_{det}| < 0.0044 \times 0.030$. As indicated earlier,
the fraction of muons from $W$ decay in this sample is negligible.

In conclusion, the results (\ref{final_A_5}) through (\ref{Afb_dimuon}),
are consistent with the standard model \cite{pdg}, and constrain
some of its extensions \cite{randall,hewett}.
The general result (\ref{Ageneral}) complements measurements
at $B$ factories, which are sensitive only to
$A_{B^0}$, not $A_{B_s^0}$.

\begin{acknowledgments}

%
We thank the staffs at Fermilab and collaborating institutions, 
and acknowledge support from the 
DOE and NSF (USA);
CEA and CNRS/IN2P3 (France);
FASI, Rosatom and RFBR (Russia);
CAPES, CNPq, FAPERJ, FAPESP and FUNDUNESP (Brazil);
DAE and DST (India);
Colciencias (Colombia);
CONACyT (Mexico);
KRF and KOSEF (Korea);
CONICET and UBACyT (Argentina);
FOM (The Netherlands);
PPARC (United Kingdom);
MSMT (Czech Republic);
CRC Program, CFI, NSERC and WestGrid Project (Canada);
BMBF and DFG (Germany);
SFI (Ireland);
The Swedish Research Council (Sweden);
Research Corporation;
Alexander von Humboldt Foundation;
and the Marie Curie Program.

\end{acknowledgments}

\end{document}